\documentclass[aps,prd,showpacs,eqsecnum,amsmath,amssymb,nofootinbib,onecolumn]{revtex4}

\usepackage[utf8]{inputenc}
\usepackage{graphicx}       
\usepackage{subfigure}
\usepackage{array}
\usepackage{multirow}
\usepackage{amsmath}
\usepackage{amssymb}
\usepackage{gensymb}
\usepackage{graphicx}
\usepackage{multirow}
\usepackage{amsmath,bm}

\usepackage{wrapfig}

\usepackage{hyperref}
\usepackage{cleveref}

\begin{document}


\title{Parameter estimation accuracies of Galactic binaries with eLISA}

\author{Arkadiusz B{\l}aut}
\address{Institute of Theoretical Physics, University of Wroc{\l}aw, Poland\\
arkadiusz.blaut@ift.uni.wroc.pl}




\begin{abstract}
We study parameter estimation accuracy of nearly monochromatic sources of gravitational waves with the future eLISA-like detectors. eLISA will be capable of observing millions of such signals generated 
by orbiting pairs of compact binaries consisting of white dwarf, neutron star or black hole and to resolve and estimate parameters of several thousands of them providing crucial information regarding their orbital dynamics, formation rates and evolutionary paths. Using the Fisher matrix analysis we compare accuracies of the estimated parameters for different mission designs defined by the GOAT advisory team established to asses the scientific capabilities and the technological issues of the eLISA-like missions.
\end{abstract}





\maketitle

\section{Introduction}
\label{intro}
The Galactic compact binary star systems are the prominent sources of gravitational waves (GW's) that will be detected by the future space--based gravitational wave interferometers. 
The LISA Pathfinder project \cite{LPF}, the important first step toward the GW space--interferometry, is already underway. 
By testing key technologies in deep space it paves the path for the large--scale missions like eLISA \cite{eLISA}, 
the most elaborated candidate, recently accepted as a part of the L$3$ Cosmic Vision 
ESA program within the theme named ``Gravitational Wave Universe`` with a possible launch in $2034$.
The most numerous systems in the eLISA band are the Galactic compact binaries containing white dwarfs (WD's), neutron stars (NS's) 
and in smaller number black holes (BH's) with the orbital periods shorter 
than a few hours. eLISA will be capable of observing millions of compact binaries and to resolve and estimate parameters of several thousands of them, 
see e.g. \cite{AS2012}, \cite{NVNP2012}, providing crucial information on the formation rates, evolutionary paths and the end products of binary systems.
Currently a several tens of optical, UV, and X-ray binaries are observed in the eLISA band
and a few of them with the orbital periods less than $15$ minutes, see e.g. \cite{Roelofs2010}. Some of those systems, qualified as {\it verification binaries}, 
are guaranteed sources for the detection. Most of the observed short--period binaries are the interacting systems, AM CVn stars, ultra--compact X-ray binaries (UCBX) 
or cataclysmic variables whose orbital evolution is influenced by a mass transfer
between components. It is noteworthy that the highest sensitivity of the eLISA-like detectors involve frequencies that are typical of the frequency--reversal transition 
of the AM CVn and UCBX stars describing the pre--contact and post--contact dynamics of those systems; for a review of the evolution of compact binary stars see \cite{PY2014}. 
On the other hand the early studies \cite{Web84} and the population synthesis models, e.g. \cite{NYPZ01}, indicate that the Galactic GW signal should be prevailed 
by the detached double white dwarf binaries (DDWD) evolving basically according to the general--relativistic radiation--reaction interaction. The loudest and the shortest--period observed detached
system, $12.75$ minute SDSS J0651+2844 white dwarf binary \cite{Kilic2011}, would be detected by eLISA already after a short period of observation.

In the present paper we study the accuracy of the parameter estimation of the Galactic compact binaries with the eLISA-like detectors. 
Since the number of the analysed systems is rather large to this end we use the Fisher information matrix (FIM) method that extensively have been used 
in previous researches e.g. \cite{Cut98}, \cite{TS2002}, \cite{AB2011} for the LISA interferometer, as well as for eLISA detector \cite{LLNC12}. 
Here we study the parameter--estimations accuracies by using the full 
response of the moving eLISA detector. We compare the distributions of errors for various detector designs
laid down by a "Gravitational Observatory Advisory Team" (GOAT) \cite{GOAT} established by ESA to asses the technological and scientific issues of the eLISA-like missions.

The organization of the paper is as follows. In Section ~\ref{sec:1} we briefly introduce the eLISA mission concept and the response of the detector to nearly monochromatic sources; 
Section ~\ref{sec:2} briefly recalls the notion of the detector sensitivity and definition of the Fisher matrix--based estimation accuracies. 
Section ~\ref{sec:3} compares the distribution of errors for different detector configurations.

\section{Detector response}
\label{sec:1}

The eLISA detector consists of three spacecraft freely orbiting the Sun in the configuration of approximately equilateral triangle. The center of mass of the constellation 
orbits the Sun at the distance $R=1$AU, the detector plane is inclined $60$ degrees with respect to the ecliptic and, seen from the Sun, the spacecraft cartwheel once per orbit. 
The final configuration of the eLISA mission is postponed to around 2020. Following the GOAT report and \cite{Klein2016} we analyse 24 mission configurations
N$i$A$j$M$k$L$l$, where $i=1,2$, $j=1,2,5$, $k=2,5$, $l=4,6$. Labels N$1$ and N$2$ denote two different levels of the acceleration noise; 
A$1$, A$2$, A$5$ stand for three arm lengths of the detector, $1$, $2$ and $5\times 10^6$km respectively; M$2$ and M$5$ define mission lifetimes of $2$ and $5$ years; 
L$4$, L$6$ indicate the use of either 4 or 6 laser links. 

The dominant laser frequency noise can be effectively reduced by the use of the time--delay interferometry (TDI) \cite{TA99}, \cite{DNV2002}. 
For the L4 design four laser links are established between the spacecraft 1 (``mother``) and the two others (''daughters''). 
The so called unequal arm Michelson $X$ TDI observable centered at the spacecraft 1 is given by
\begin{eqnarray}
\label{eq:X}
X(t) & = & y_{12}(t-4L) + y_{21}(t-3L) + y_{13}(t-L) + y_{31}(t)  \\
&& -\left[ y_{13}(t-4L) + y_{31}(t-3L) + y_{12}(t-L) + y_{21}(t) \right].
\end{eqnarray}
In the above expression $y_{ab}(t)$ is the one-way GW--induced laser frequency shift $\frac{\Delta\nu}{\nu}$ measured along the arm $L{\bf n}_{ab}$, 
where ${\bf n}_{ab}$ is the unit vector oriented from the emitter $a$ to the receiver $b$ and $L$ is the length of the link. For a plane GW 
${\bf h}(t-{\bf k}\cdot{\bf x})$ with the wave--propagation unit vector ${\bf k}$ the response at the detection point $(t,{\bf x}_{b})$ reads \cite{EW75}
\begin{eqnarray}
\label{eq:y}
y_{ab}(t) & = & \frac{{\bf n}_{ab}\otimes{\bf n}_{ab}}{2(1-{\bf k}\cdot{\bf n}_{ab})}:
\left[ {\bf h}(t-L-{\bf k}\cdot{\bf x}_a) - {\bf h}(t-{\bf k}\cdot{\bf x}_b) \right],
\end{eqnarray}
where $a,b = 1,2,3$ and '$:$' denotes contraction. For a nearly monochromatic signal
\begin{eqnarray}
\label{eq:wave}
{\bf h}({\bf x},t) & = & h^{+}\cos{\left[\tilde\omega(t) 
\cdot (t-{\bf k}\cdot{\bf x})+\phi_0\right]}{\bm\epsilon}^{+} +
h^{\times}\sin{\left[\tilde\omega(t) \cdot (t-{\bf k}\cdot{\bf x})+\phi_0\right]}{\bm\epsilon}^{\times},
\end{eqnarray}
with the two polarization tensors $\bm{\epsilon}^{+}$ and $\bm\epsilon^{\times}$
and the time--dependent angular frequency $\tilde\omega(t) = \omega + \frac12\dot{\omega}t + \frac16\ddot{\omega}t^2$ the basic response takes the form
\begin{eqnarray}
\label{eq:yab}
y_{ab}(t) & = &
\omega L\,\text{sinc}\left[\frac{\omega L}{2}\left(1-{\bf k}\cdot{\bf n}_{ab}\right)\right]\,
\left[ a_1\,u_{ab}(t) \sin{\Phi_{ab}(t) + a_2\,v_{ab}(t)} \sin{\Phi_{ab}(t)}\;+\right.\nonumber\\
&&\quad\quad\quad\quad\quad\quad\quad\quad
\left. - \;a_3\,u_{ab}(t) \cos{\Phi_{ab}(t)} - a_4\,v_{ab}(t) \cos{\Phi_{ab}(t)} \right]\\
\label{eq:yab_Phi}
\Phi_{ab}(t) & = & \tilde{\omega}(t)t - \sigma(t){\bf k}\cdot{\bf x}_{b} -
\frac{\omega L}{2}\left(1-{\bf k}\cdot{\bf n}_{ab}\right),\qquad
\sigma(t)  =  \omega + \dot{\omega}t + \frac12\ddot{\omega}t^2.
\end{eqnarray}
In Eqs. (\ref{eq:yab}) and (\ref{eq:yab_Phi}) $\omega$, $\dot{\omega}$ and $\ddot{\omega}$ denote time--independent quantities determined at the initial time of the observation $t=0$. 
The four time--independent amplitudes $a_i$ are functions of the astrophysical parameters of the binary, the strain amplitude $h$, 
the inclination of the orbit $\iota$, the polarization angle $\psi$, and the initial phase $\phi_0$ of the signal. The explicit form of the amplitudes $a_i$ can be found e.g. in \cite{KTV04};
the two polarization amplitudes $h^{+}$ and $h^{\times}$ are functions of $h$ and $\iota$,
\begin{eqnarray}
\label{eq:polampl}
h^{+} = \frac{h}{2} (1+\cos^2{\iota}),\qquad
h^{\times} = h \cos{\iota};
\end{eqnarray}
for slowly evolving detached binaries the amplitude $h$ reads
\begin{eqnarray}
\label{eq:ampl}
h = \frac{4(G{\cal M}_c)^{5/3}}{c^4\,D}\left[\frac{\omega}{2}\right]^{2/3},
\end{eqnarray}
where $D$ is the luminosity distance to the source and
${\cal M}_c=m_1^{3/5}m_2^{3/5}/(m_1+m_2)^{1/5}$ is the {\it chirp mass}. For the noninteracting systems the chirping angular frequency is given by
\begin{eqnarray}
\label{eq:angfr}
\dot{\omega} = \frac{48}{5}\left(\frac{G{\cal M}_c}{2\,c^3}\right)^{5/3}\omega^{11/3}, \qquad
\ddot{\omega} = \frac{11}{3}\frac{\dot{\omega}^2}{\omega}.
\end{eqnarray}
The {\it amplitude modulation} functions $u_{ab}(t)$ and $v_{ab}(t)$ are defined by the decomposition \cite{KTV04}:
\begin{eqnarray}
\label{eq:uv}
F^{+}\left[{\bf n}_{ab};t\right] =
u_{ab}(t)\cos{2\psi} + v_{ab}(t)\sin{2\psi},\\
F^{\times}\left[{\bf n}_{ab}(t)\right] =
v_{ab}(t)\cos{2\psi} - u_{ab}(t)\sin{2\psi},
\end{eqnarray}
where $F^{\pi}\left[{\bf n}_{ab}\right]=\frac12{\boldsymbol\epsilon}^{\pi}:{\bf n}_{ab}\otimes{\bf n}_{ab}$
for $\pi=+,\;\times$. The non-zero components of the polarization tensors in the source frame are  
$\epsilon^{+}_{11}=-\epsilon^{+}_{22}=1$ and $\epsilon^{\times}_{12}=\epsilon^{\times}_{21}=1$.

\begin{figure}[htp]
\begin{center}
\includegraphics[width=20pc]{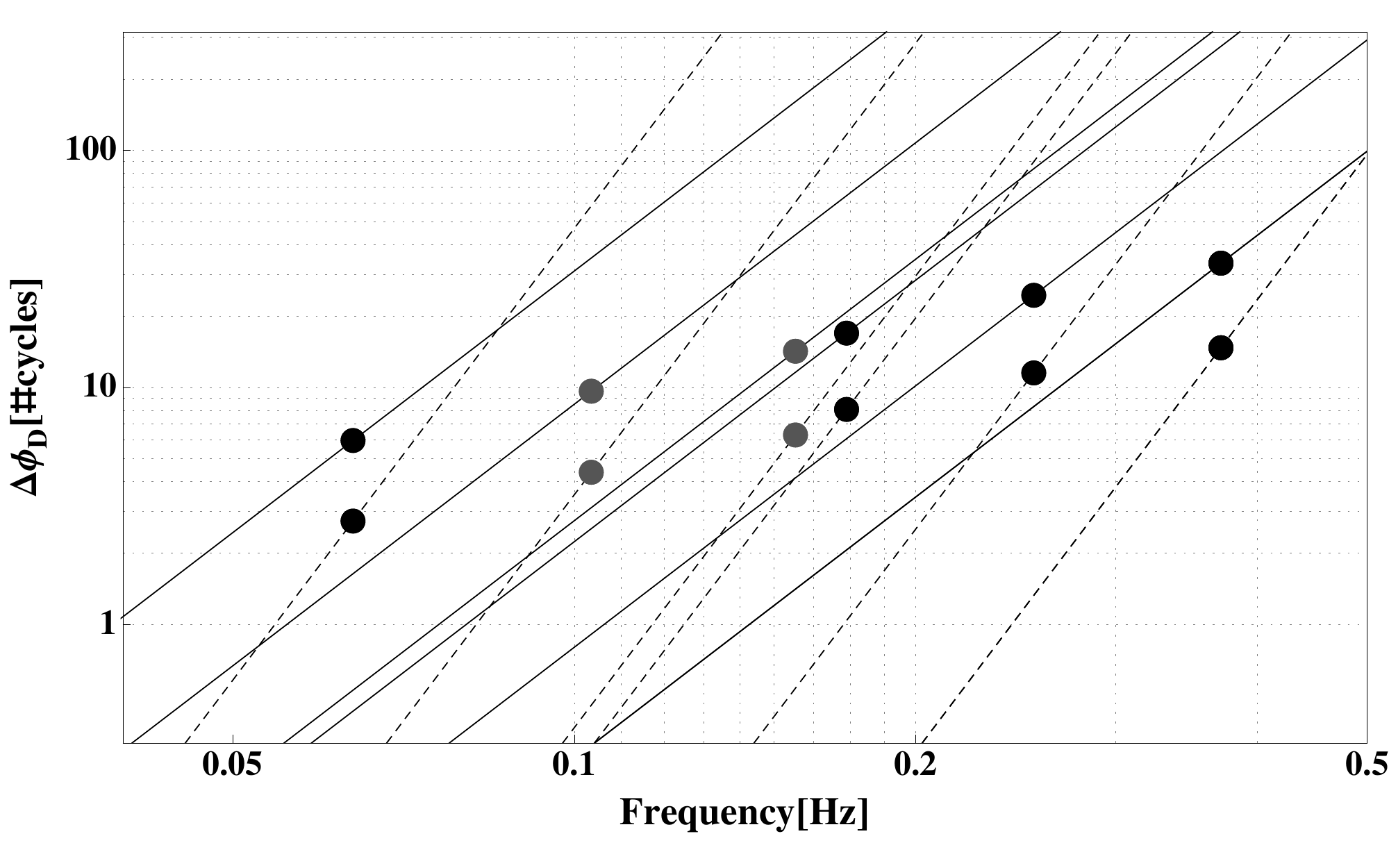}
\end{center}
\caption{Contributions of the Doppler terms, $\frac12\dot{\omega}t_{obs}R$ (continuous) and
$\frac16\ddot{\omega}t^2_{obs}R$ (dashed) to the phase $\Phi_{ab}$ expressed as a number of cycles in two years. 
Filled circles denote binaries with 
the coalescence time strictly equal to the time of observation ($2$ years); from the right: 
WD--WD, WD--NS, WD--BH, NS-NS, NS--BH, BH--BH for $0.35M_\odot$WD, $1.4M_\odot$NS and $6M_\odot$BH.}
\label{fig:Doppler_phase}
\end{figure}

For the equilateral triangular configuration of the spacecraft it is convenient to separate the guiding center ${\bf x}_{gc}$ holding the ecliptic plane 
and to decompose the position vectors into ${\bf x}_b={\bf x}_{gc}+{\bf p}_b$, where ${\bf p}_b$ is the vector from the guiding center to the vertex at 
${\bf x}_b$ \cite{KTV04}. The delayed response $y_{ab}(t-kL)$ then has the form of Eq. (\ref{eq:yab}) with the phase $\Phi_{ab}$ given by
\begin{eqnarray}
\label{eq:Phi}
\Phi_{ab}(t-kL) = \tilde{\omega}(t)t - \sigma(t){\bf k}\cdot{\bf x}_{gc} -
\left(k + \frac12 + d_c\right)\,\omega L,\qquad k=1,2,\ldots
\end{eqnarray}
where index c completes $a,b$ in the set $\{1,2,3\}$ and 
$d_1=\frac{{\bf n}_{13}-{\bf n}_{21}}{6}\cdot{\bf k}$, $d_2=\frac{{\bf n}_{21}-{\bf n}_{32}}{6}\cdot{\bf k}$, $d_3=\frac{{\bf n}_{32}-{\bf n}_{13}}{6}\cdot{\bf k}$. 
In the time-delayed response $y_{ab}(t-kL)$ the retardation $t-kL$ is ignored in the slowly varying functions of time $u$, $v$ and ${\bf n}_{ab}$, and the terms 
$\dot{\omega}t_{obs}L$, $\ddot{\omega}t^2_{obs}L$ are neglected. The Solar System Barycenter (SSB) coordinates are used in which $|{\bf x}_{gc}|=R\simeq500$s. 
In what follows we shall keep terms containing frequency rates, Eq. (\ref{eq:angfr}), in the phase although their contributions to the Doppler modulation terms, 
$\dot{\omega}\,{\bf x}_{g}(t)$ and $\ddot{\omega}\,{\bf x}_{g}(t)$, are non-negligible only above $f=\frac{\omega}{2\pi}\sim40$mHz, see Fig. \ref{fig:Doppler_phase}.

The total noise consists of the instrumental noise and the foreground noise from the unresolvable Galactic white--dwarf binaries \cite{NVNP2012}. 
The instrumental noise spectral density for the $X$ observable is given by
\begin{eqnarray}
\label{eq:noise_i}
S_{X,i} = 16\sin^2{x}\left[ 2\left( 1+\cos^2{x} \right)S^{pm}_{\frac{\delta\nu}{\nu}} + S^{op}_{\frac{\delta\nu}{\nu}} \right],\qquad x=2\pi f L
\end{eqnarray}
where $S^{op}_{\frac{\delta\nu}{\nu}}$ is the spectral density due to the shot noise and other kinds of optical noises and $S^{pm}_{\frac{\delta\nu}{\nu}}$ is due to the acceleration noise. 
For the strain Galactic confusion noise $S_{h,gal}$ we take the analytical expressions developed in \cite{Klein2016} and \cite{Timpano2006}; 
explicit forms of the noises are given in the \ref{A1}. The total noise $S_{X}$ is the sum of the $S_{X,i}$ and the sky-averaged X-equivalent Galactic noise
\begin{eqnarray}
\label{eq:noise_G}
S_{X,gal} = \frac{3}{20}\frac{16 x^2 \sin^2{x}}{1 + \left(\frac{x}{0.41\pi}\right)^2}\times S_{h,gal}.
\end{eqnarray}

The six laser links offer the opportunity of using three interferometric observables for which the noises are uncorrelated. 
For the L6 design we take $A=(Z-X)/\sqrt{2}$, $E=(X-2Y+Z)/\sqrt{6}$, $T=(X+Y+Z)/\sqrt{3}$ combinations, where $Y$ and $Z$ are the unequal arm Michelson observables 
centered at the spacecraft 2 and 3.

\section{Sensitivity}
\label{sec:2}

\begin{figure}[htp]
\begin{center}
\hspace{-1.cm} \includegraphics[width=30pc]{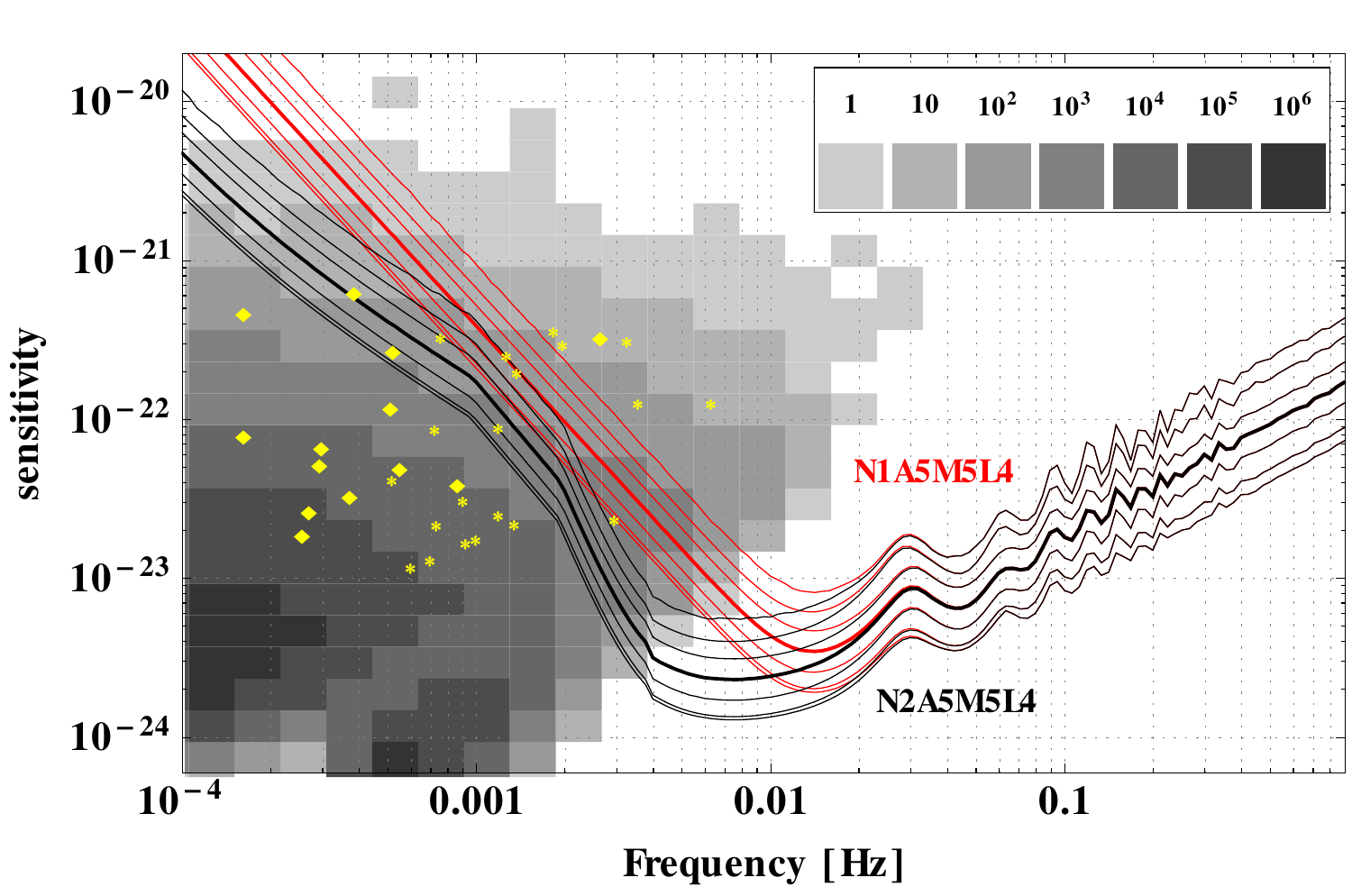}
\end{center}
\caption{Sensitivity curves for the randomly oriented Galactic binaries; N1A5M5L4, N2A5M5L4 configurations. 
The continuous lines denote $0.1\%$, $3\%$, $25\%$, $50\%$ (thick curve), $75\%$, $97\%$ and $99.9\%$ quantiles. 
Gray boxes give the amplitude number density for the population synthesis model \cite{NYPZ01}. Some known detached (diamonds) and AM CVn (stars) verification binaries are depicted.}
\label{fig:fmap:sensi1}
\end{figure}

\begin{figure}[htp]
\begin{center}
\hspace{-1.cm} \includegraphics[width=30pc]{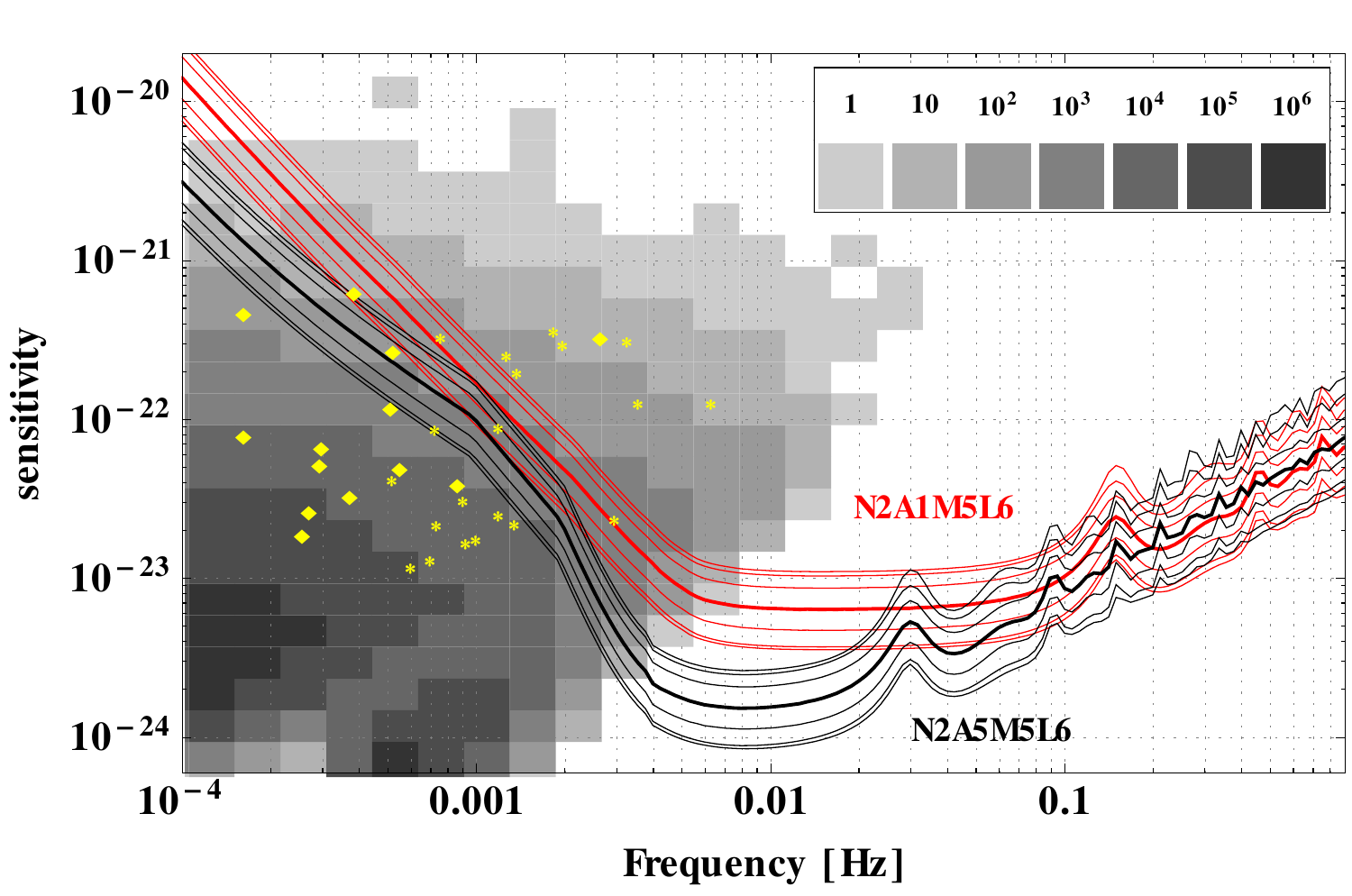}
\end{center}
\caption{Sensitivity curves for the Galactic, randomly oriented binaries; N2A1M5L6, N2A5M5L6 configurations.}
\label{fig:fmap:sensi2}
\end{figure}

The optimal signal to noise ratio $\rho_{\text{opt}}$ for a nearly monochromatic signal can be approximated by the square root of $\frac{2}{S_{X}(\omega)}\int\limits_0^{t_{obs}}X^2(t)dt$, 
where $t_{obs}$ is the observation time; the sensitivity for a detection threshold $\rho_{\text{thr}}$ is then defined as
$\frac{\rho_{\text{thr}}}{\rho_{\text{opt}}(h=1)}$.
Figs. \ref{fig:fmap:sensi1} and \ref{fig:fmap:sensi2} depict sensitivities as a function of GW frequency for some chosen detector configurations with assumed threshold $\rho_{\text{thr}}=5$. 
The continuous lines represent $0.1\%$, $3\%$, $25\%$, $50\%$, $75\%$, $97\%$ and $99.9\%$ quantiles of the frequency--dependent distributions obtained 
in the Monte Carlo simulations over a $10^5$ binaries. Randomly oriented sources were distributed according to the thick--disk Galaxy model \cite{NYPZ01} with the probability density given by
\begin{eqnarray}
\label{eq:thick_Galaxy}
\frac{dN}{drdz} \sim e^{-r/r_0}\text{sech}^2(z/z_0);\qquad\text{$r_0=2.5$kpc and $z_0=200$pc},
\end{eqnarray}
where $r$ and $z$ are the cylindrical coordinates, the distance from the Galactic center and the hight above the Galactic plane; 
simulations were performed in frequency intervals of the length $\Delta\ln{f}=0.025$.
The strip of $0.1\%$--$99.9\%$ quantiles can be used to represent non--sky--averaged detection threshold for the Galactic binaries. 
It provides an information about the fraction of signals with different amplitudes that are detected with the signal to noise ratio $\rho=5$; cf. \cite{MV08}. 
In Figs. \ref{fig:fmap:sensi1} and \ref{fig:fmap:sensi2} two pairs of configurations, N1A5M5L4, N2A5M5L4 and N2A1M5L6, N2A5M5L6 were chosen to illustrate 
the influence of the acceleration noise level  (represented by the parameter N) and the length of the arm (represented by the parameter A) on the detector sensitivity.

The gray boxes shown in Figs. \ref{fig:fmap:sensi1} and \ref{fig:fmap:sensi2} give the amplitude number densities for the population of $\sim4.5\times 10^7$ 
detached and semi--detached systems taken from the population synthesis model \cite{NYPZ01} used in the series of MLDC challenges \cite{MLDCs}. 
In addition a sample of detached (diamonds) and AM CVn (stars) systems observed optically in the detectors band is depicted. 
The loudest of them are potential candidates for the detection and can serve as a verification binaries; their parameters are taken from \cite{vb}. 
The shortest--period systems are, among the non--interacting systems, the $2.61$mHz  J$0651$ detached binary and $6.22$mHz HMCnC for AM CVn stars. 
\begin{figure}[htp]
\begin{center}
\hspace{0.cm} \includegraphics[width=29pc]{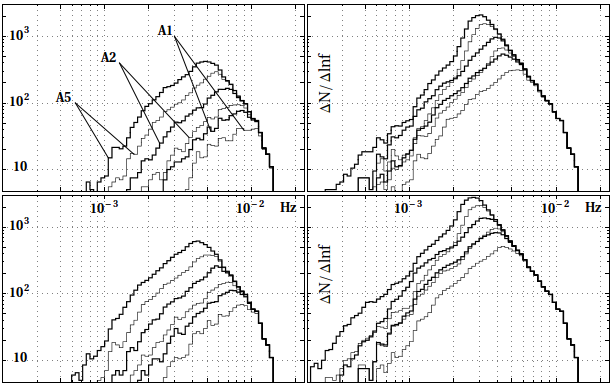}
\end{center}
\caption{Number density $\Delta N(f)/\Delta\ln{f}$ of binaries above the median of the sensitivity ($\rho_{\text{thr}}=5$) as a function of the frequency; $\Delta\ln{f}=0.025$. 
Upper panel: L4, lower panel: L6; left column: N1, right column: N2; thin lines: M2, thick lines: M5.}
\label{fig:fmap:counts}
\end{figure}

Fig. \ref{fig:fmap:counts} shows the number of detectable Galactic sources as a function of frequency for all detector configurations. 
The curves present the number densities $\frac{\Delta N}{\Delta\ln{f}}$ ($\Delta\ln{f}=0.025$) of the population of Galactic binaries above the median of the $\rho=5$ detection threshold. 
The improvement of the detector design (lower acceleration noise, longer arm, longer mission lifetime, more links) results in an increase of the number of the detectable sources. 
Fig. \ref{fig:fmap:counts} indicates that that number is mostly affected by the acceleration noise (N) which determines the width of the sensitivity band for low frequencies 
where the majority of the population resides. The plots indicate that in the case of the N2 option A1M5 and A2M2 configurations follow the same pattern; 
A5M2 is better (worse) than A2M5 above (below) $2$mHz. For the N1 option longer arm improves the detection capabilities independently on the mission lifetime.

\section{Parameters}
\label{sec:3}

To analyse the errors we use the Fisher information matrix which for the nearly monochromatic signals can be approximated by
\begin{eqnarray}
\label{eq:fim}
\Gamma_{ij} = \frac{2}{S_{X}(\omega)}\int\limits_0^{t_{obs}}\frac{\partial X}{\partial \theta^i}\frac{\partial X}{\partial \theta^j}dt,
\end{eqnarray}
where $\theta^i$ are parameters to be estimated in some detection procedure. 
According to the Cramer-Rao theorem, the errors defined by $\sigma_{\theta_i}=\sqrt{C_{ii}}$ , where
$C=\Gamma^{-1}$, define the lower bounds among all non--biased estimators.
We take $8$-dimensional parameter space including $f$, $\dot{f}$, ecliptic latitude $\beta$ and longitude $\lambda$ of the source and the astrophysical parameters; 
in the case of the second time derivative of the frequency the $9$-dimensional parameter space is considered. The angular resolutions of the sky position 
and orientation of the angular momentum of the orbit are given by $\Delta\Omega=2\pi\cos{\beta}\sqrt{C_{\beta\beta}C_{\lambda\lambda}-C_{\beta\lambda}^2}$ and
$\Delta\Omega_L=2\pi\sin{\iota}\sqrt{C_{\iota\iota}C_{\psi\psi}-C_{\psi\iota}^2}$ respectively. The resolution angle $\Delta\Omega_L$ is defined with respect 
to the sky--location--dependent frame; $z$--axis of this frame is the propagation unit vector, $x$--axis is tangent to the lines of constant $\lambda$ 
and is directed toward northern ecliptic hemisphere.

\subsection{Distribution of errors}
\label{sec:31}

\begin{figure}[htp]
\begin{center}
%
\includegraphics[width=32pc]{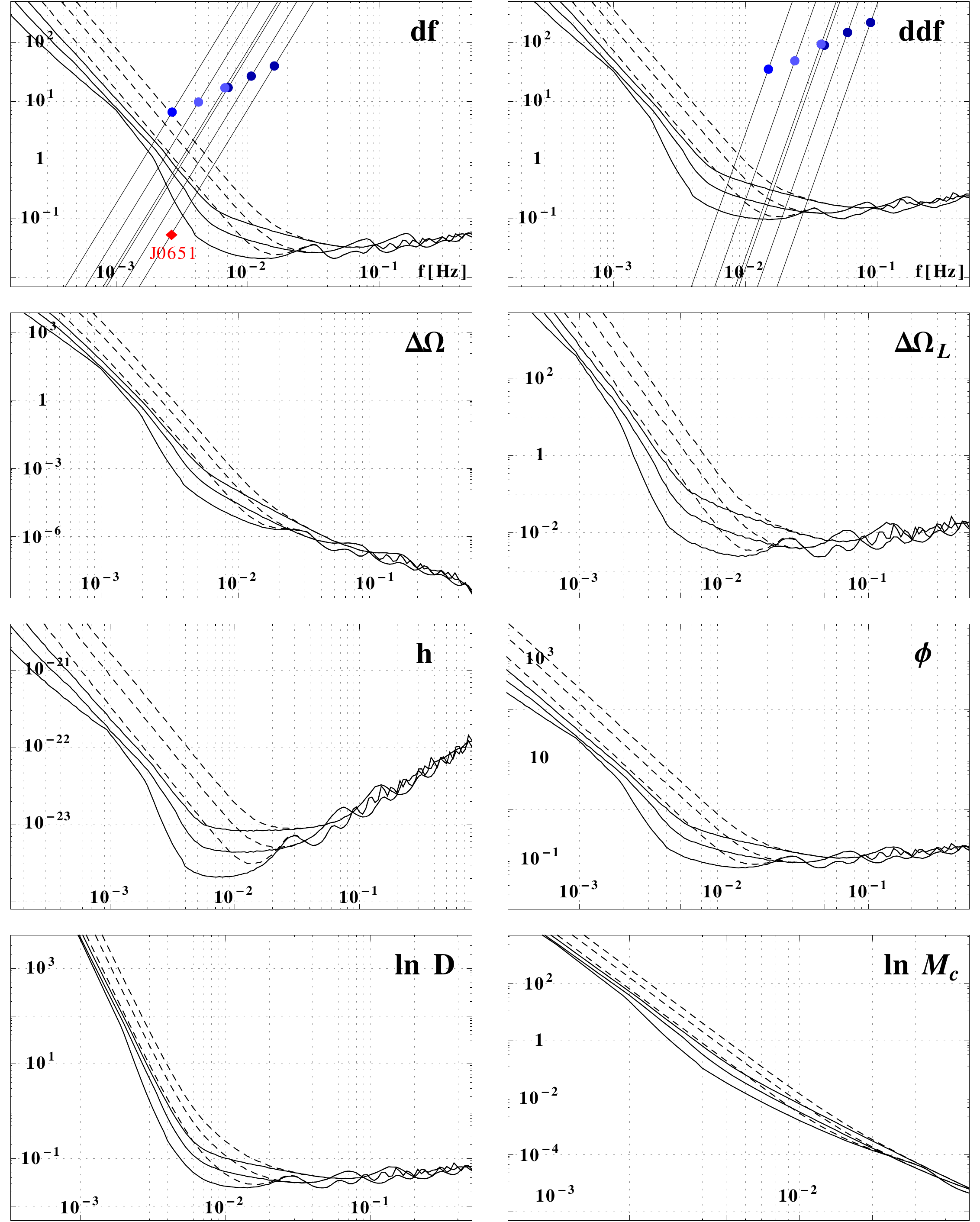}
\end{center}
\caption{
Medians of the errors for different M2L4 missions as functions of the GW frequency: N1A1, N1A2, N1A5 (upper, middle, lower dashed lines) and N2A1, N2A2, N2A5 (upper, middle, lower solid lines).
The frequency rate errors are scaled to give the number of cycles accumulated by the corresponding terms in the signal phase during the observation time. 
The straight dashed lines denote the total number of cycles for those terms. The filled circles stand for the systems with the coalescence time equal to $10^4$ years (df) and $10^2$ years (ddf) 
for a six types of binaries consisting of $m=0.35M_{\odot}$ DDWD, $m=1.4M_{\odot}$ NS and $m=6M_{\odot}$ BH; from the right: WD--WD, WD--NS, WD--BH, NS-NS, NS--BH, BH--BH. 
The small diamond denotes the location of J0651 detached binary.
}
\label{fig:vars_f}
\end{figure}

To investigate the frequency dependence of the errors we apply the Monte Carlo simulations of 
equal--mass ($m_1=m_2=0.35$M$_{\odot}$), randomly oriented Galactic binaries. Fig. \ref{fig:vars_f} shows the medians of the errors as functions of the GW frequency. 
The extension of the mission lifetime from two (M2) to five (M5) years increases the signal to noise ratio by the factor $\sqrt{5/2}$. 
As a consequence this improves the parameter estimation accuracies by the factor $\sqrt{5/2}$ (this is also the case for the frequency and frequency rates errors 
if they are expressed in terms of the number of cycles; see below) and resolution angles approximately by $2/5$. In turn the signal to noise ratio improvement 
between six link (L6) and four link (L4) configurations gives factors $\sqrt{2}$ in the long wavelength limit (below frequencies $~1/L$) and $\sqrt{3}$, 
on average, for higher frequencies \cite{PTL02}. Therefore in Fig. 3 we show parameter estimation errors only for the four--link, two-years missions. 

The frequency rate errors, $\sigma_{\text{df}}$ and $\sigma_{\text{ddf}}$, are expressed in terms of $\;\frac12\sigma_{\dot{f}}t^2_{obs}$ and $\;\frac16\sigma_{\ddot{f}}t^3_{obs}$. 
This gives the numbers of cycles due to errors in the corresponding Doppler terms in the phase (\ref{eq:Phi}) accumulated during the observation time. 
The straight dashed lines give the total number of cycles due to $\;\frac122\pi\dot{f}t^2_{obs}$ and $\;\frac162\pi\ddot{f}t^3_{obs}$ terms for a six types of sources 
consisting of pairs of $m=0.35M_{\odot}$ white dwarfs (WD), $m=1.45M_{\odot}$ neutron stars (NS) and $m=6M_{\odot}$ black holes (BH); the filled circles on the diagrams 
identify the coalescence time equal to $10^4$ years  (for df) and $10^2$ years  (for ddf) for each type of the binary. The numbers of cycles due to the errors compared 
with the total number of cycles can give a hint on when the first and second frequency derivatives for detached binaries ought to be included in the detector response in a data analysis.

Diagrams in Fig. \ref{fig:vars_f} reveal the following features. The highest accuracies are obtained for the N2A5 mission in the whole frequency band; 
below $5$mHz all depicted N2 configurations are more favourable than N1; N1A5 configuration has better parameter estimation capabilities than N2A1 and N2A2 for frequencies 
greater than $6$mHz and $10$mHz respectively. 
These medium and high frequency properties are interesting for systems revealing the frequency–-reversal transition (AM CVn, UCBX-types of stars) describing the pre-–contact 
and post-–contact dynamics of those systems.
It may be worth recalling here that for interacting systems mass transfer starts at frequencies $1$--$2$mHz for binaries 
with helium star component, at $~20$mHz for double $0.35M_{\odot}$ white dwarfs, and for more massive double WD's or for WD's with NS or BH companions at higher frequencies \cite{Nel03}. 
Binaries consisting of two NS's or a pairs of NS and BH's start to interact at frequencies above $\sim 1$kHz outside the eLISA band. 
The expected number of systems with neutron stars or black holes at high frequency eLISA band is however negligible or highly uncertain \cite{NYPZ01}, \cite{BelBul10}.

Estimation of the distance and the chirp mass of the radiating detached binary is achievable when the amplitude and the frequency derivative $\dot{f}$ can be estimated; 
from Eq. (\ref{eq:ampl}) and from the first equation of (\ref{eq:angfr}) it then follows
\begin{eqnarray}
\label{eq:DM}
D\sim \frac{\dot{f}}{h\,f^3},\qquad {\cal M}_c\sim \frac{\dot{f}^{3/5}}{f^{11/5}}.
\end{eqnarray}
In both cases the contribution of the frequency errors is negligible as compared to the contribution of the
errors due to the amplitude and frequency rate. Therefore for the chirp mass the approximation 
$\sigma_{\ln{{\cal M}_c}}\simeq\frac35\sigma_{\ln\dot{f}}$ is valid in the whole band.
For the luminosity distance $D$ there are two regimes: at low frequencies the distance is determined by the
frequency derivative $\sigma_{\ln{D}}\simeq\sigma_{\ln\dot{f}}$, above $\sim (5-10)$mHz it is determined by
the amplitude, $\sigma_{\ln{D}}\simeq\sigma_{\ln{h}}$. The transitions between the two regimes is only slightly configuration--dependent.

\begin{figure}[htp]
\begin{center}
\includegraphics[width=28pc]{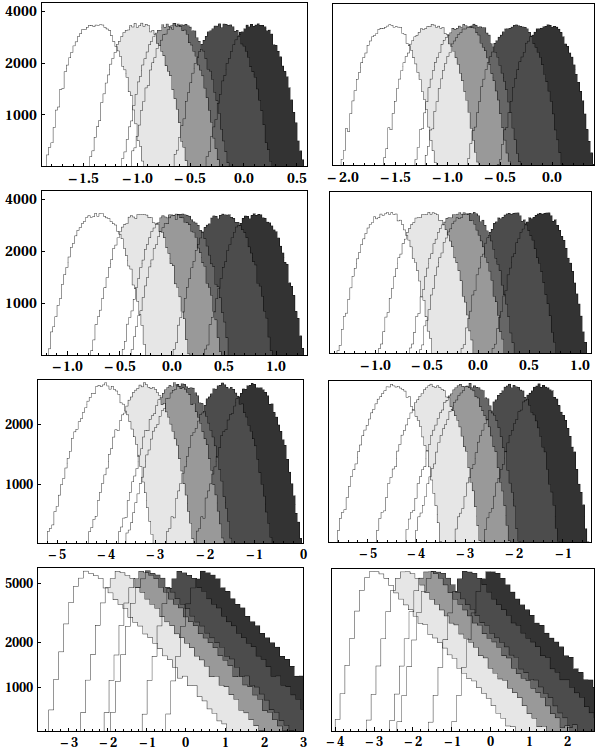}
\end{center}
\caption{
Distribution of errors for $10^5$ equal--mass ($m_1=m_2=0.35$M${}_{\odot}$), randomly oriented Galactic binaries in the $\Delta\ln{f}=0.0125$ frequency interval at $5$mHz.
From above: $\sigma_{df}$, $\sigma_{ddf}$, $\Delta\Omega$, $\Delta\Omega_L$; left column: M2L4 configuration; right column: M2L6 configuration; grey scale corresponds to N2A5(white), N2A2, N2A1, N1A5, N1A2, N1A1(black) mission designs. For all errors log${}_{10}$ scale is used.
}
\label{fig:vars_1}
\end{figure}

\begin{figure}[htp]
\begin{center}
\includegraphics[width=28pc]{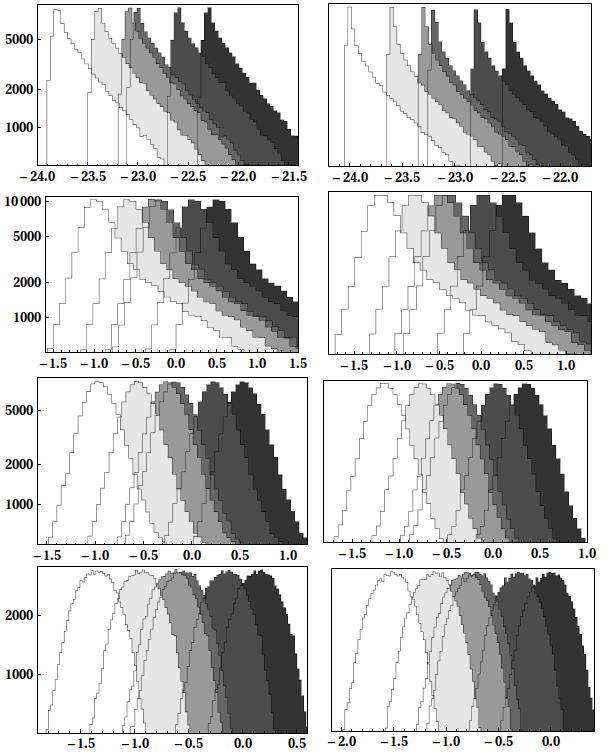}
\end{center}
\caption{Distribution of errors at $5$mHz; from above: $\sigma_{h}$, $\sigma_{\phi}$, $\sigma_{ln\,D}$, $\sigma_{\ln\,M}$. Conventions as in Fig. \ref{fig:vars_1}.
}
\label{fig:vars_2}
\end{figure}

Estimation errors at fixed frequency of $5$mHz are presented in Figs. \ref{fig:vars_1} and \ref{fig:vars_2}. 
Distributions of $\sigma_h$, $\sigma_\phi$, $\Delta\Omega_L$ reveal an asymmetry which arises due to strong dependence of the astrophysical parameters errors, 
$\sigma_h$, $\sigma_\phi$ $\sigma_\iota$, $\sigma_\psi$, on the source orientation.
The most probable values of those errors are lower than the medians depicted on Fig. \ref{fig:vars_f}. 
The same behaviour become apparent for the luminosity distance parameter $D$ starting at frequencies where
the $\sigma_{\ln{D}}$ errors are determined by the amplitude $h$.

\begin{figure}[htp]
\begin{center}
\includegraphics[width=30pc]{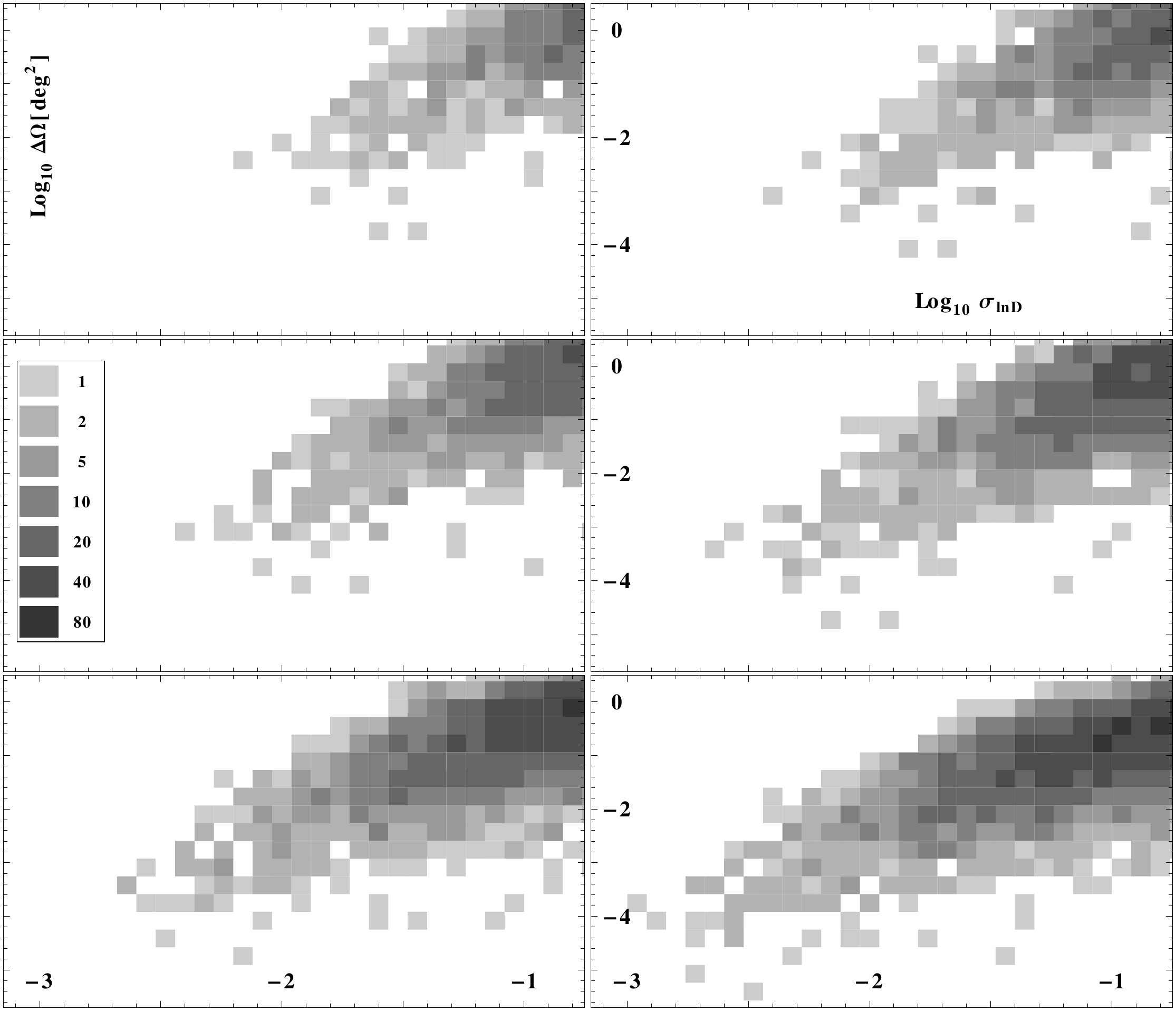}
\end{center}
\caption{$\sigma_{\ln\,D}-\Delta\Omega$ diagrams for N2M2 configurations; left column: L4, right column: L6;
from above: A1, A2, A5.}
\label{fig:domM2}
\end{figure}

\begin{figure}[htp]
\begin{center}
\includegraphics[width=30pc]{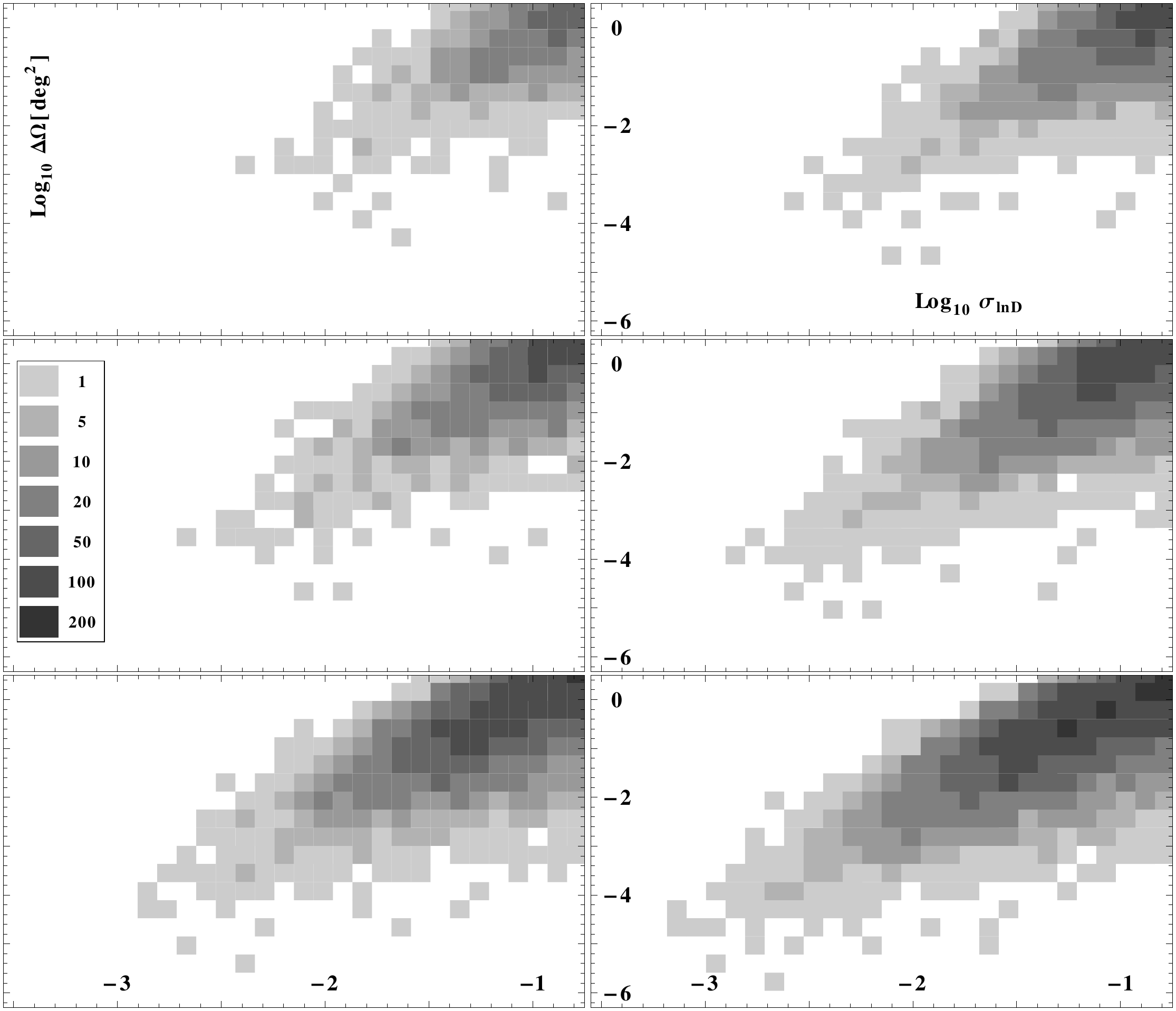}
\end{center}
\caption{$\sigma_{\ln\,D}-\Delta\Omega$ diagrams for N2M5 configurations. Conventions as in Fig. \ref{fig:domM2}.}
\label{fig:domM5}
\end{figure}

Figs. \ref{fig:domM2} and \ref{fig:domM5} demonstrate the capability of localizing Galactic binaries by the detectors for the all N2 configurations. 
On the distance-angular resolution diagrams the background gray boxes show the number densities of the detached double white dwarf Galactic binaries 
taken from the population synthesis model of \cite{NYPZ01}; selected sources have signal to noise ratio $\rho\geq 7$.

\begin{figure}[htp]
\begin{center}
\includegraphics[width=25pc]{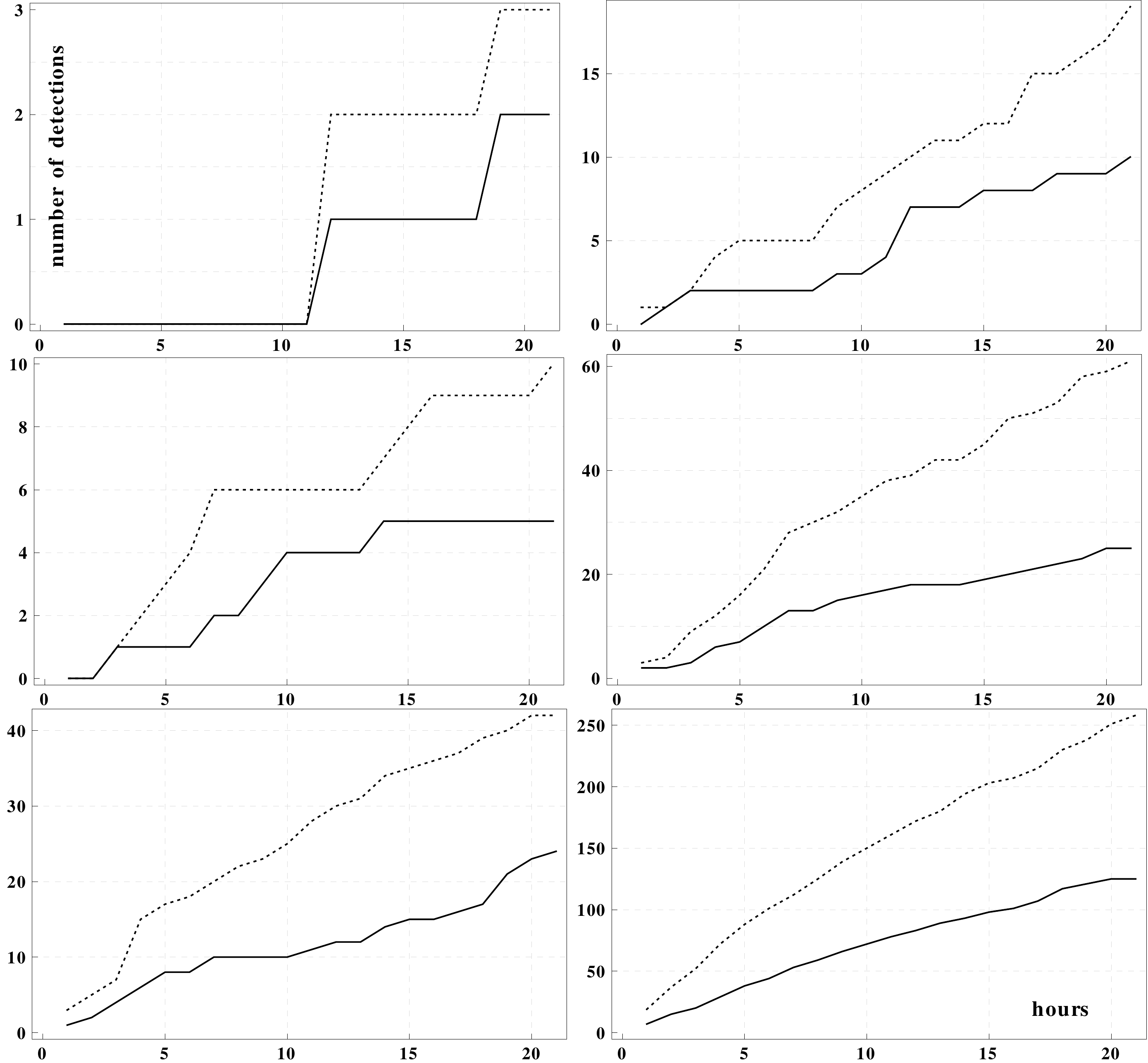}
\end{center}
\caption{Number of detected signals ($\text{snr }\geq 7$) during first 20 hours; columns: N1-left, N2-right, rows: A1-upper, A2-middle, A5-lower; L4-solid line, L6-dashed line.}
\label{fig:nrt_time_7}
\end{figure}

Tab. \ref{t:DOmega} gives total numbers of sources with $\rho\geq 7$, $\Delta\Omega<\pi$[deg${}^2$] and 
$\sigma_{\ln\,D}<0.1$ for all mission designs.
\begin{table}
\begin{center}
\caption{Number of binaries ploted on Figs. \ref{fig:domM2}, \ref{fig:domM5} for all detector configurations; 
selected are signals with  $\rho\geq 7$, $\Delta\Omega<\pi$[deg${}^2$] and 
$\sigma_{\ln\,D}<0.1$.}
\renewcommand{\arraystretch}{1.3}
\begin{tabular}{c c c c c c c c c}
\hline\hline
&
\multicolumn{1}{p{1.cm}}
{\centering \footnotesize{  } }
&
\multicolumn{1}{p{1.cm}}
{\centering \footnotesize{  } }
&
\multicolumn{1}{p{1.cm}}
{\centering \footnotesize{ A1M2 } }
&
\multicolumn{1}{p{1.cm}}
{\centering \footnotesize{ A2M2 } }
&
\multicolumn{1}{p{1.cm}}
{\centering \footnotesize{ A5M2 } }
&
\multicolumn{1}{p{1.cm}}
{\centering \footnotesize{ A1M5 } }
&
\multicolumn{1}{p{1.cm}}
{\centering \footnotesize{ A2M5 } }
&
\multicolumn{1}{p{1.cm}}
{\centering \footnotesize{ A5M5 } }
\\
\hline
& \multicolumn{1}{p{1.cm}}
{\centering \footnotesize{ L4 } }
& \multicolumn{1}{p{1.cm}}
{\centering \footnotesize{ N1 } }
& \multicolumn{1}{p{1.cm}}
{\centering \footnotesize{ $43$ } }
& \multicolumn{1}{p{1.cm}}
{\centering \footnotesize{ $88$ } }
& \multicolumn{1}{p{1.cm}}
{\centering \footnotesize{ $111$ } }
& \multicolumn{1}{p{1.cm}}
{\centering \footnotesize{ $216$ } }
& \multicolumn{1}{p{1.cm}}
{\centering \footnotesize{ $340$ } }
& \multicolumn{1}{p{1.cm}}
{\centering \footnotesize{ $732$ } }
\\
& \multicolumn{1}{p{1.cm}}
{\centering \footnotesize{  } }
& \multicolumn{1}{p{1.cm}}
{\centering \footnotesize{ N2 } }
& \multicolumn{1}{p{1.cm}}
{\centering \footnotesize{ $229$ } }
& \multicolumn{1}{p{1.cm}}
{\centering \footnotesize{ $616$ } }
& \multicolumn{1}{p{1.cm}}
{\centering \footnotesize{ $608$ } }
& \multicolumn{1}{p{1.cm}}
{\centering \footnotesize{ $1512$ } }
& \multicolumn{1}{p{1.cm}}
{\centering \footnotesize{ $1557$ } }
& \multicolumn{1}{p{1.cm}}
{\centering \footnotesize{ $4267$ } }
\\
\hline
& \multicolumn{1}{p{1.cm}}
{\centering \footnotesize{ L6 } }
& \multicolumn{1}{p{1.cm}}
{\centering \footnotesize{ N1 } }
& \multicolumn{1}{p{1.cm}}
{\centering \footnotesize{ $86$ } }
& \multicolumn{1}{p{1.cm}}
{\centering \footnotesize{ $160$ } }
& \multicolumn{1}{p{1.cm}}
{\centering \footnotesize{ $205$ } }
& \multicolumn{1}{p{1.cm}}
{\centering \footnotesize{ $415$ } }
& \multicolumn{1}{p{1.cm}}
{\centering \footnotesize{ $619$ } }
& \multicolumn{1}{p{1.cm}}
{\centering \footnotesize{ $1270$ } }
\\
& \multicolumn{1}{p{1.cm}}
{\centering \footnotesize{  } }
& \multicolumn{1}{p{1.cm}}
{\centering \footnotesize{ N2 } }
& \multicolumn{1}{p{1.cm}}
{\centering \footnotesize{ $477$ } }
& \multicolumn{1}{p{1.cm}}
{\centering \footnotesize{ $1299$ } }
& \multicolumn{1}{p{1.cm}}
{\centering \footnotesize{ $1048$ } }
& \multicolumn{1}{p{1.cm}}
{\centering \footnotesize{ $2868$ } }
& \multicolumn{1}{p{1.cm}}
{\centering \footnotesize{ $2328$ } }
& \multicolumn{1}{p{1.cm}}
{\centering \footnotesize{ $6303$ } }
\\
\hline
\end{tabular}
\label{t:DOmega}
\end{center}
\end{table}
Tabs. \ref{t:D2D3_N1} and \ref{t:D2D3_N2} show how the detection and estimation capabilities depend on the time of observation.
They give the number of detected sources (with $\rho\geq 7$), the number of sources well localized in the sky ($\Delta\Omega\leq 1$deg${}^2$), 
number of sources well localized in the 3--dimensional volume ($\Delta\Omega\leq 1$deg${}^2$ and the distance determined to better than $10\%$), and the number of sources 
having chirp mass estimated to better than $20\%$; the results are given for all L6 missions. They can be compared to some recent studies. E.g. \cite{CR17}
considers the two Michelson channels (corresponding to L6 configuration), and the $2.5$ million kilometer--long arms (close to but longer than in A2 configuration).
The subtraction procedure used in \cite{CR17} strongly influences modeling of the confusion noise and effectively lowers the Galactic foreground in a time--dependent way. 
Thus although one can observe a similar short observation time results (compared with A2N2 configuration of the present paper) the long time behaviour is different: 
subtraction procedure inreases the number of detected sources gradually (up to 20-40\% for 4--year observation period).
\begin{table}
\begin{center}
\caption{Number of binaries for N1M5L6 detector configurations detected in 6months, 1 year, 2 years, 4 years, and 5 years; 
selected are signals with  $\text{snr }\geq 7$ (\# detected), $\Delta\Omega<1$[deg${}^2$] (2D mapped),
$\Delta\Omega<1$[deg${}^2$]$, \sigma_{\ln\,D}<0.1$ (3D mapped), and  $\sigma_{\ln\,M}<0.2$ (${\cal M}_c$ measured).}
\renewcommand{\arraystretch}{1.3}
\begin{tabular}{c c c c c c c c}
\hline\hline
&
\multicolumn{1}{p{2.cm}}
{\centering \footnotesize{  } }
&
\multicolumn{1}{p{1.cm}}
{\centering \footnotesize{  } }
&
\multicolumn{1}{p{1.3cm}}
{\centering \footnotesize{ 6 months } }
&
\multicolumn{1}{p{1.3cm}}
{\centering \footnotesize{ 1 year } }
&
\multicolumn{1}{p{1.3cm}}
{\centering \footnotesize{ 2 years } }
&
\multicolumn{1}{p{1.3cm}}
{\centering \footnotesize{ 4 years } }
&
\multicolumn{1}{p{1.3cm}}
{\centering \footnotesize{ 5 years } }
\\
\hline
& \multicolumn{1}{p{2.cm}}
{\centering \footnotesize{  } }
& \multicolumn{1}{p{1.cm}}
{\centering \footnotesize{ A1 } }
& \multicolumn{1}{p{1.3cm}}
{\centering \footnotesize{ $374$ } }
& \multicolumn{1}{p{1.3cm}}
{\centering \footnotesize{ $613$ } }
& \multicolumn{1}{p{1.3cm}}
{\centering \footnotesize{ $946$ } }
& \multicolumn{1}{p{1.3cm}}
{\centering \footnotesize{ $1442$ } }
& \multicolumn{1}{p{1.3cm}}
{\centering \footnotesize{ $1635$ } }
\\
& \multicolumn{1}{p{2.cm}}
{\centering \footnotesize{ \# detected } }
& \multicolumn{1}{p{1.cm}}
{\centering \footnotesize{ A2 } }
& \multicolumn{1}{p{1.3cm}}
{\centering \footnotesize{ $917$ } }
& \multicolumn{1}{p{1.3cm}}
{\centering \footnotesize{ $1447$ } }
& \multicolumn{1}{p{1.3cm}}
{\centering \footnotesize{ $2130$ } }
& \multicolumn{1}{p{1.3cm}}
{\centering \footnotesize{ $3155$ } }
& \multicolumn{1}{p{1.3cm}}
{\centering \footnotesize{ $3570$ } }
\\
& \multicolumn{1}{p{2.cm}}
{\centering \footnotesize{  } }
& \multicolumn{1}{p{1.cm}}
{\centering \footnotesize{ A5 } }
& \multicolumn{1}{p{1.3cm}}
{\centering \footnotesize{ $2683$ } }
& \multicolumn{1}{p{1.3cm}}
{\centering \footnotesize{ $4073$ } }
& \multicolumn{1}{p{1.3cm}}
{\centering \footnotesize{ $5823$ } }
& \multicolumn{1}{p{1.3cm}}
{\centering \footnotesize{ $8307$ } }
& \multicolumn{1}{p{1.3cm}}
{\centering \footnotesize{ $9314$ } }
\\
\hline
& \multicolumn{1}{p{2.cm}}
{\centering \footnotesize{  } }
& \multicolumn{1}{p{1.cm}}
{\centering \footnotesize{ A1 } }
& \multicolumn{1}{p{1.3cm}}
{\centering \footnotesize{ $9$ } }
& \multicolumn{1}{p{1.3cm}}
{\centering \footnotesize{ $71$ } }
& \multicolumn{1}{p{1.3cm}}
{\centering \footnotesize{ $280$ } }
& \multicolumn{1}{p{1.3cm}}
{\centering \footnotesize{ $419$ } }
& \multicolumn{1}{p{1.3cm}}
{\centering \footnotesize{ $479$ } }
\\
& \multicolumn{1}{p{2.cm}}
{\centering \footnotesize{ 2D mapped } }
& \multicolumn{1}{p{1.cm}}
{\centering \footnotesize{ A2 } }
& \multicolumn{1}{p{1.3cm}}
{\centering \footnotesize{ $19$ } }
& \multicolumn{1}{p{1.3cm}}
{\centering \footnotesize{ $151$ } }
& \multicolumn{1}{p{1.3cm}}
{\centering \footnotesize{ $603$ } }
& \multicolumn{1}{p{1.3cm}}
{\centering \footnotesize{ $869$ } }
& \multicolumn{1}{p{1.3cm}}
{\centering \footnotesize{ $962$ } }
\\
& \multicolumn{1}{p{2.cm}}
{\centering \footnotesize{  } }
& \multicolumn{1}{p{1.cm}}
{\centering \footnotesize{ A5 } }
& \multicolumn{1}{p{1.3cm}}
{\centering \footnotesize{ $65$ } }
& \multicolumn{1}{p{1.3cm}}
{\centering \footnotesize{ $444$ } }
& \multicolumn{1}{p{1.3cm}}
{\centering \footnotesize{ $1374$ } }
& \multicolumn{1}{p{1.3cm}}
{\centering \footnotesize{ $1953$ } }
& \multicolumn{1}{p{1.3cm}}
{\centering \footnotesize{ $2159$ } }
\\
\hline
& \multicolumn{1}{p{2.cm}}
{\centering \footnotesize{  } }
& \multicolumn{1}{p{1.cm}}
{\centering \footnotesize{ A1 } }
& \multicolumn{1}{p{1.3cm}}
{\centering \footnotesize{ $4$ } }
& \multicolumn{1}{p{1.3cm}}
{\centering \footnotesize{ $20$ } }
& \multicolumn{1}{p{1.3cm}}
{\centering \footnotesize{ $83$ } }
& \multicolumn{1}{p{1.3cm}}
{\centering \footnotesize{ $132$ } }
& \multicolumn{1}{p{1.3cm}}
{\centering \footnotesize{ $153$ } }
\\
& \multicolumn{1}{p{2.cm}}
{\centering \footnotesize{ 3D mapped } }
& \multicolumn{1}{p{1.cm}}
{\centering \footnotesize{ A2 } }
& \multicolumn{1}{p{1.3cm}}
{\centering \footnotesize{ $10$ } }
& \multicolumn{1}{p{1.3cm}}
{\centering \footnotesize{ $43$ } }
& \multicolumn{1}{p{1.3cm}}
{\centering \footnotesize{ $185$ } }
& \multicolumn{1}{p{1.3cm}}
{\centering \footnotesize{ $317$ } }
& \multicolumn{1}{p{1.3cm}}
{\centering \footnotesize{ $378$ } }
\\
& \multicolumn{1}{p{2.cm}}
{\centering \footnotesize{  } }
& \multicolumn{1}{p{1.cm}}
{\centering \footnotesize{ A5 } }
& \multicolumn{1}{p{1.3cm}}
{\centering \footnotesize{ $30$ } }
& \multicolumn{1}{p{1.3cm}}
{\centering \footnotesize{ $132$ } }
& \multicolumn{1}{p{1.3cm}}
{\centering \footnotesize{ $558$ } }
& \multicolumn{1}{p{1.3cm}}
{\centering \footnotesize{ $962$ } }
& \multicolumn{1}{p{1.3cm}}
{\centering \footnotesize{ $1130$ } }
\\
\hline
& \multicolumn{1}{p{2.cm}}
{\centering \footnotesize{  } }
& \multicolumn{1}{p{1.cm}}
{\centering \footnotesize{ A1 } }
& \multicolumn{1}{p{1.3cm}}
{\centering \footnotesize{ $56$ } }
& \multicolumn{1}{p{1.3cm}}
{\centering \footnotesize{ $152$ } }
& \multicolumn{1}{p{1.3cm}}
{\centering \footnotesize{ $994$ } }
& \multicolumn{1}{p{1.3cm}}
{\centering \footnotesize{ $2508$ } }
& \multicolumn{1}{p{1.3cm}}
{\centering \footnotesize{ $3299$ } }
\\
& \multicolumn{1}{p{2.cm}}
{\centering \footnotesize{ ${\cal M}_c$ measured } }
& \multicolumn{1}{p{1.cm}}
{\centering \footnotesize{ A2 } }
& \multicolumn{1}{p{1.3cm}}
{\centering \footnotesize{ $91$ } }
& \multicolumn{1}{p{1.3cm}}
{\centering \footnotesize{ $248$ } }
& \multicolumn{1}{p{1.3cm}}
{\centering \footnotesize{ $1425$ } }
& \multicolumn{1}{p{1.3cm}}
{\centering \footnotesize{ $3475$ } }
& \multicolumn{1}{p{1.3cm}}
{\centering \footnotesize{ $4456$ } }
\\
& \multicolumn{1}{p{2.cm}}
{\centering \footnotesize{  } }
& \multicolumn{1}{p{1.cm}}
{\centering \footnotesize{ A5 } }
& \multicolumn{1}{p{1.3cm}}
{\centering \footnotesize{ $184$ } }
& \multicolumn{1}{p{1.3cm}}
{\centering \footnotesize{ $457$ } }
& \multicolumn{1}{p{1.3cm}}
{\centering \footnotesize{ $2264$ } }
& \multicolumn{1}{p{1.3cm}}
{\centering \footnotesize{ $5254$ } }
& \multicolumn{1}{p{1.3cm}}
{\centering \footnotesize{ $6778$ } }
\\
\hline
\end{tabular}
\label{t:D2D3_N1}
\end{center}
\end{table}
\begin{table}
\begin{center}
\caption{Number of binaries for N2M5L6 configurations. Conventions as in Tab.\ref{t:D2D3_N1}.}
\renewcommand{\arraystretch}{1.3}
\begin{tabular}{c c c c c c c c}
\hline\hline
&
\multicolumn{1}{p{2.cm}}
{\centering \footnotesize{  } }
&
\multicolumn{1}{p{1.cm}}
{\centering \footnotesize{  } }
&
\multicolumn{1}{p{1.3cm}}
{\centering \footnotesize{ 6 months } }
&
\multicolumn{1}{p{1.3cm}}
{\centering \footnotesize{ 1 year } }
&
\multicolumn{1}{p{1.3cm}}
{\centering \footnotesize{ 2 years } }
&
\multicolumn{1}{p{1.3cm}}
{\centering \footnotesize{ 4 years } }
&
\multicolumn{1}{p{1.3cm}}
{\centering \footnotesize{ 5 years } }
\\
\hline
& \multicolumn{1}{p{2.cm}}
{\centering \footnotesize{  } }
& \multicolumn{1}{p{1.cm}}
{\centering \footnotesize{ A1 } }
& \multicolumn{1}{p{1.3cm}}
{\centering \footnotesize{ $3381$ } }
& \multicolumn{1}{p{1.3cm}}
{\centering \footnotesize{ $5482$ } }
& \multicolumn{1}{p{1.3cm}}
{\centering \footnotesize{ $8181$ } }
& \multicolumn{1}{p{1.3cm}}
{\centering \footnotesize{ $11855$ } }
& \multicolumn{1}{p{1.3cm}}
{\centering \footnotesize{ $13344$ } }
\\
& \multicolumn{1}{p{2.cm}}
{\centering \footnotesize{ \# detected } }
& \multicolumn{1}{p{1.cm}}
{\centering \footnotesize{ A2 } }
& \multicolumn{1}{p{1.3cm}}
{\centering \footnotesize{ $6477$ } }
& \multicolumn{1}{p{1.3cm}}
{\centering \footnotesize{ $9456$ } }
& \multicolumn{1}{p{1.3cm}}
{\centering \footnotesize{ $13252$ } }
& \multicolumn{1}{p{1.3cm}}
{\centering \footnotesize{ $18471$ } }
& \multicolumn{1}{p{1.3cm}}
{\centering \footnotesize{ $20615$ } }
\\
& \multicolumn{1}{p{2.cm}}
{\centering \footnotesize{  } }
& \multicolumn{1}{p{1.cm}}
{\centering \footnotesize{ A5 } }
& \multicolumn{1}{p{1.3cm}}
{\centering \footnotesize{ $13609$ } }
& \multicolumn{1}{p{1.3cm}}
{\centering \footnotesize{ $18094$ } }
& \multicolumn{1}{p{1.3cm}}
{\centering \footnotesize{ $23060$ } }
& \multicolumn{1}{p{1.3cm}}
{\centering \footnotesize{ $29568$ } }
& \multicolumn{1}{p{1.3cm}}
{\centering \footnotesize{ $30539$ } }
\\
\hline
& \multicolumn{1}{p{2.cm}}
{\centering \footnotesize{  } }
& \multicolumn{1}{p{1.cm}}
{\centering \footnotesize{ A1 } }
& \multicolumn{1}{p{1.3cm}}
{\centering \footnotesize{ $33$ } }
& \multicolumn{1}{p{1.3cm}}
{\centering \footnotesize{ $311$ } }
& \multicolumn{1}{p{1.3cm}}
{\centering \footnotesize{ $1438$ } }
& \multicolumn{1}{p{1.3cm}}
{\centering \footnotesize{ $2185$ } }
& \multicolumn{1}{p{1.3cm}}
{\centering \footnotesize{ $2436$ } }
\\
& \multicolumn{1}{p{2.cm}}
{\centering \footnotesize{ 2D mapped } }
& \multicolumn{1}{p{1.cm}}
{\centering \footnotesize{ A2 } }
& \multicolumn{1}{p{1.3cm}}
{\centering \footnotesize{ $98$ } }
& \multicolumn{1}{p{1.3cm}}
{\centering \footnotesize{ $762$ } }
& \multicolumn{1}{p{1.3cm}}
{\centering \footnotesize{ $2741$ } }
& \multicolumn{1}{p{1.3cm}}
{\centering \footnotesize{ $3830$ } }
& \multicolumn{1}{p{1.3cm}}
{\centering \footnotesize{ $4202$ } }
\\
& \multicolumn{1}{p{2.cm}}
{\centering \footnotesize{  } }
& \multicolumn{1}{p{1.cm}}
{\centering \footnotesize{ A5 } }
& \multicolumn{1}{p{1.3cm}}
{\centering \footnotesize{ $379$ } }
& \multicolumn{1}{p{1.3cm}}
{\centering \footnotesize{ $2181$ } }
& \multicolumn{1}{p{1.3cm}}
{\centering \footnotesize{ $6097$ } }
& \multicolumn{1}{p{1.3cm}}
{\centering \footnotesize{ $8014$ } }
& \multicolumn{1}{p{1.3cm}}
{\centering \footnotesize{ $8654$ } }
\\
\hline
& \multicolumn{1}{p{2.cm}}
{\centering \footnotesize{  } }
& \multicolumn{1}{p{1.cm}}
{\centering \footnotesize{ A1 } }
& \multicolumn{1}{p{1.3cm}}
{\centering \footnotesize{ $13$ } }
& \multicolumn{1}{p{1.3cm}}
{\centering \footnotesize{ $68$ } }
& \multicolumn{1}{p{1.3cm}}
{\centering \footnotesize{ $424$ } }
& \multicolumn{1}{p{1.3cm}}
{\centering \footnotesize{ $895$ } }
& \multicolumn{1}{p{1.3cm}}
{\centering \footnotesize{ $1088$ } }
\\
& \multicolumn{1}{p{2.cm}}
{\centering \footnotesize{ 3D mapped } }
& \multicolumn{1}{p{1.cm}}
{\centering \footnotesize{ A2 } }
& \multicolumn{1}{p{1.3cm}}
{\centering \footnotesize{ $27$ } }
& \multicolumn{1}{p{1.3cm}}
{\centering \footnotesize{ $165$ } }
& \multicolumn{1}{p{1.3cm}}
{\centering \footnotesize{ $969$ } }
& \multicolumn{1}{p{1.3cm}}
{\centering \footnotesize{ $2062$ } }
& \multicolumn{1}{p{1.3cm}}
{\centering \footnotesize{ $2435$ } }
\\
& \multicolumn{1}{p{2.cm}}
{\centering \footnotesize{  } }
& \multicolumn{1}{p{1.cm}}
{\centering \footnotesize{ A5 } }
& \multicolumn{1}{p{1.3cm}}
{\centering \footnotesize{ $94$ } }
& \multicolumn{1}{p{1.3cm}}
{\centering \footnotesize{ $367$ } }
& \multicolumn{1}{p{1.3cm}}
{\centering \footnotesize{ $2316$ } }
& \multicolumn{1}{p{1.3cm}}
{\centering \footnotesize{ $4892$ } }
& \multicolumn{1}{p{1.3cm}}
{\centering \footnotesize{ $5720$ } }
\\
\hline
& \multicolumn{1}{p{2.cm}}
{\centering \footnotesize{  } }
& \multicolumn{1}{p{1.cm}}
{\centering \footnotesize{ A1 } }
& \multicolumn{1}{p{1.3cm}}
{\centering \footnotesize{ $123$ } }
& \multicolumn{1}{p{1.3cm}}
{\centering \footnotesize{ $356$ } }
& \multicolumn{1}{p{1.3cm}}
{\centering \footnotesize{ $2461$ } }
& \multicolumn{1}{p{1.3cm}}
{\centering \footnotesize{ $6037$ } }
& \multicolumn{1}{p{1.3cm}}
{\centering \footnotesize{ $7853$ } }
\\
& \multicolumn{1}{p{2.cm}}
{\centering \footnotesize{ ${\cal M}_c$ measured } }
& \multicolumn{1}{p{1.cm}}
{\centering \footnotesize{ A2 } }
& \multicolumn{1}{p{1.3cm}}
{\centering \footnotesize{ $219$ } }
& \multicolumn{1}{p{1.3cm}}
{\centering \footnotesize{ $563$ } }
& \multicolumn{1}{p{1.3cm}}
{\centering \footnotesize{ $3405$ } }
& \multicolumn{1}{p{1.3cm}}
{\centering \footnotesize{ $7674$ } }
& \multicolumn{1}{p{1.3cm}}
{\centering \footnotesize{ $9693$ } }
\\
& \multicolumn{1}{p{2.cm}}
{\centering \footnotesize{  } }
& \multicolumn{1}{p{1.cm}}
{\centering \footnotesize{ A5 } }
& \multicolumn{1}{p{1.3cm}}
{\centering \footnotesize{ $415$ } }
& \multicolumn{1}{p{1.3cm}}
{\centering \footnotesize{ $1001$ } }
& \multicolumn{1}{p{1.3cm}}
{\centering \footnotesize{ $5297$ } }
& \multicolumn{1}{p{1.3cm}}
{\centering \footnotesize{ $11255$ } }
& \multicolumn{1}{p{1.3cm}}
{\centering \footnotesize{ $13791$ } }
\\
\hline
\end{tabular}
\label{t:D2D3_N2}
\end{center}
\end{table}
Fig. \ref{fig:nrt_time_7}, in addition , shows the early time detection capabilities for all missions.

\section{Summary}
\label{sec:4}

We have studied the accuracy of the parameters estimation of nearly monochromatic GWs signals from Galactic binaries with the LISA-like detectors using the time delay interferometry. 
We have compared detection capabilities of different mission designs proposed by the GOAT advisory team. 

To compare sensitivities we have performed in each frequency interval Monte Carlo simulation of the thick Galaxy model.
The diagrams \ref{fig:fmap:sensi1}, \ref{fig:fmap:sensi2} reveal the impact of the acceleration noise level and the detector arm length on the sesitivity.
Results are presented in the non--sky--averaged form which enables an estimation of the overlap between sensitivity strips of different configurations.
In addition we show on these diagrams the amplitude number densities of Galactic white dwarf binaries population developed in the population synthesis model \cite{NYPZ01}.
Related to theses diagrams is the number density $\Delta N(f)/\Delta\ln f$ plot of the detectable sources given in Fig. \ref{fig:fmap:counts}.
It confirms the conclusion of the advisory team \cite{GOAT} on the strong impact of the acceleration noise. It also indicates that in the case of N1 option 
(for both of 4 and 6--link configuration) the quality of detection improves in the following order: A1M2, A1M5, A2M2, A2M5, A5M2, A5M5; 
the same ramains valid for the N2 option with the exeption that A1M5 exceed A2M2 below 3mHz and A2M5 exceeds A5M2 below 2mHz. 
This is related to the fact that the N2 missions dig more deeply into the Galactic foreground in which case the longer mission 
lifetme is more effective than the longer lenght of the detector's arm. 

Parameter estimation accuracies, likewise, have been studied in the non--sky--averaged form by performing in each frequency interval Monte Carlo simulations 
of mass-normalized Galactic sources. The results are presented in Fig. \ref{fig:vars_f}. It gives frequency dependence of the medians of the distributions and compares the influence of 
the acceleration noise level and the detector arm length on the estimated parameters. 
Figs. \ref{fig:vars_1}, \ref{fig:vars_2} show distributions of all errors at 5mHz and in addition demonstrate how the number of links improves the estimation. 
We can observe and assess the strong impact of the arm length and the acceleration noise level and the comparatively moderate influence of the number of links on the estimated parameters.
It is worth pointing out that similar (but not presented in the paper) diagrams comparing influence of the arm length and the mission lifetime on the accuracy of estimated parameters
give the same  A1M2, A1M5, A2M2, A2M5, A5M2, A5M5 order for most parameters (except for the mass and the distance in which case the mission lifetime is more decisive). 
Again, due to the Galactic confusion noise A1M5 exceed A2M2 and A2M5 exceeds A5M2 below ~2--3mHz those parameters for missions with lower acceleration noise. 
This also indicates that further studies taking into account succesive subtraction scheme of the foreground noise for all considered as well as other potential configurations would be appropriate. 
In this regard some results obtained in the present paper have been compared with the recent studies \cite{CR17} and from this point of view interpreted.

Finally, 3D--localization capabilities were demonstrated on the distance--angular resolution diagrams for all N2 configurations, 
and the expected number of the early time detections were presented for all missions.

\section{Acknowledgements}
I would like to thank Andrzej Kr\'olak for helpful discussions.
The work was supported in part by the National Science Centre of Poland Grant UMO-2014/14/M/ST9/00707.

\appendix

\section{Instrumental and Galactic confusion noises}

The following values of the of the instrumental noises are considered (in [Hz${}^{-1}$]):

\[
S^{pm}_{\frac{\delta\nu}{\nu}}(f) = 
	\begin{cases}
	9\times 10^{-28}\frac{1}{(2\pi f c)^2}\left( 1 + \frac{10^{-4}\text{Hz}}{f} \right) & \text{for N}1 \\
	9\times 10^{-30}\frac{1}{(2\pi f c)^2}\left( 1 + \frac{10^{-4}\text{Hz}}{f} \right) & \text{for N}2  
	\end{cases}
\]

\[
S^{op}_{\frac{\delta\nu}{\nu}}(f) = 
	\begin{cases}
	(2.65 + 1.98)\times 10^{-23}\left(\frac{2\pi f}{c}\right)^2 & \text{for A}1 \\
	(2.65 + 2.22)\times 10^{-23}\left(\frac{2\pi f}{c}\right)^2 & \text{for A}2 \\
	(2.65 + 2.96)\times 10^{-23}\left(\frac{2\pi f}{c}\right)^2 & \text{for A}3
	\end{cases}
\]

The Galactic confusion noises read:

\[ 
S_{gal,\text{N2A1}} = 
  \begin{cases}
    f^{-2.1}\times 1.155206\times 10^{-43} & \quad 10^{-5}\leq f < 5.3\times 10^{-4}\\
    f^{-3.235}\times 2.9714\times 10^{-47} & \quad 5.3\times 10^{-4}\leq f < 2.2\times 10^{-3}\\
    f^{-4.85}\times 1.517\times 10^{-51} & \quad 2.2\times 10^{-3}\leq f < 4\times 10^{-3}\\
    f^{-7.5}\times 6.706\times 10^{-58} & \quad 4\times 10^{-3}\leq f < 5.3\times 10^{-3}\\
    f^{-20.0}\times 2.39835\times 10^{-86} & \quad 5.3\times 10^{-3}\leq f \leq \times 10^{-2}
  \end{cases}
\]
\[
S_{gal,\text{N2A2}} = 
  \begin{cases}
    f^{-2.1}\times 1.3516\times 10^{-43} & \quad 10^{-5}\leq f < 5.01\times 10^{-4}\\
    f^{-3.3}\times 1.4813\times 10^{-47} & \quad 5.01\times 10^{-4}\leq f < 2.07\times 10^{-3}\\
    f^{-5.2}\times 1.17757\times 10^{-52} & \quad 2.07\times 10^{-3}\leq f < 3.4\times 10^{-3}\\
    f^{-9.1}\times 2.7781\times 10^{-62} & \quad 3.4\times 10^{-3}\leq f < 5.2\times 10^{-3}\\
    f^{-20.0}\times 3.5333\times 10^{-87} & \quad 5.2\times 10^{-3}\leq f \leq \times 10^{-2}
  \end{cases}
\]
\[
S_{gal,\text{N2A5}} = \frac{20}{3} 
  \begin{cases}
    f^{-2.3}\times 10^{-44.62} & \quad 10^{-5}\leq f < 10^{-3}\\
    f^{-4.4}\times 10^{-50.92} & \quad 10^{-3}\leq f < 10^{-2.7}\\
    f^{-8.8}\times 10^{-62.8} & \quad 10^{-2.7}\leq f < 10^{-2.4}\\
    f^{-20.0}\times 10^{-89.68} & \quad 10^{-2.4}\leq f \leq \times 10^{-2}
  \end{cases}
\]

\label{A1}

\section*{References}


\begin{thebibliography}{}
%
%
\bibitem{LPF} \url{http://sci.esa.int/lisa-pathfinder/}
%
\bibitem{eLISA} \url{https://lisamission.org.}
%
\bibitem{AS2012} P. Amaro--Seoane et al., GW notes {\bf 6} (2012).
%
\bibitem{NVNP2012} S. Nissanke, M. Vallisneri, G. Nelemans, T. Prince, ApJ  (2012).
%
\bibitem{Roelofs2010} G. H. A. Roelofs, A. Rau, T. R. Marsh et al., ApJ {\bf 711}, L$138$ (2010).
%
\bibitem{PY2014} K. A. Postnov, L. R. Yungelson, {\it Living Rev. Relativity}, {\bf 17} (2014), 3.
%
\bibitem{Web84} R. Webbink, ApJ {\bf 277}, 355 (1984).
%
\bibitem{MV08} M. Vallisneri, Phys.Rev. {\bf D} 77, 042001 (2008).
%
\bibitem{NYPZ01} G. Nelemans, L. Yungelson, S. F. Portegies--Zwart, A\&A {\bf 375}, 890 (2001).
%
\bibitem{Kilic2011} M. Kilic, W. R. Brown, S. J. Kenyon, et al., MNRAS {\bf 413}, L$101$ (2011).
%
\bibitem{Cut98} C. Cutler, Phys. Rev. {\bf D} 57, 7089 (1998).
%
\bibitem{TS2002} R. Takashi, N. Seto, ApJ {\bf 575} , 1030 (2002).
%
\bibitem{AB2011} A. B\l aut, Phys. Rev. {\bf D} 83, 083006 (2011).
%
\bibitem{LLNC12} T. B. Littenberg, S. L. Larson, G. Nelemans, N. J. Cornish, MNRAS {\bf 429} (3), 2361 (2013).
%
\bibitem{GOAT} http://www.cosmos.esa.int/web/goat
%
\bibitem{Klein2016} A. Klein {it et al.}, Phys. Rev. {\bf D} 93, 024003 (2016).
%
%
\bibitem{TA99} M. Tinto, J. W. Armstrong, Phys. Rev. {\bf D} 59, 102003 (1999).
%
\bibitem{DNV2002} S. V. Dhurandhar, K. R. Nayak, J.Y. Vinet, Phys. Rev. {\bf D} 65, 102002 (2002).
%
\bibitem{EW75} F. B. Estabrook and H. D. Wahlquist General Relativity and Gravitation {\bf 6} 439 (1975).
%
\bibitem{KTV04} A.  Kr\'olak, M. Tinto, M. Vallisneri Phys. Rev. {\bf D} 70, 022003 (2004).
%
\bibitem{Timpano2006} S. E. Timpano, L. J. Rubbo, N. J. Cornish, Phys. Rev. {\bf D} 73, 122001 (2016).
%
\bibitem{MLDCs} K. A. Arnaud {\it et al.} Class. Quant. Grav. {\bf 24} S551-S554 (2007).
%
\bibitem{vb} www.verification\_binaries.
%
\bibitem{Nel03} G. Nelemans, arXiv:astro-ph/0310800.
%
\bibitem{BelBul10} K. Belczynski, M. Benaquista, T. Bulik,  Astrophys. J. {\bf 725} 816–823 (2010).
%
\bibitem{PTL02} T. A. Prince, M. Tinto, S. L. Larson, Phys. Rev. {\bf D} 66, 122002 (2002).
%
\bibitem{CR17} N. Cornish, T. Robson, Journal of Physics: Conference Series, Volume {\bf 840}, conference 1.
%
\end{thebibliography}

\end{document}